\documentclass[conference]{IEEEtran}

\ifCLASSINFOpdf

\else

\fi

\usepackage{graphicx}
\usepackage{ifpdf}
\ifpdf
  \DeclareGraphicsExtensions{.pdf,.png,.jpg}
\else
  \DeclareGraphicsExtensions{.eps}
\fi
\usepackage{amsmath}
\usepackage{array}
\usepackage{tabularx} 
\usepackage{graphicx}
\usepackage{verbatim}
\usepackage{booktabs}
\usepackage{hyperref}
\usepackage{multirow}

\begin{document}

\title{Advanced Scheduling of Electrolyzer Modules for Grid Flexibility
}

\author{Angelina~Lesniak,
        Andrea~Gloppen~Johnsen, Noah~Rhodes, Line~Roald
        
\thanks{Department of Electrical and Computer Engineering, University of Wisconsin-Madison, Supported by a generous gift from The University of Wisconsin-Madison Graduate School and grants from the National Science Foundation (DMR-1720415), (EFMA-2132036-), (CHE-1949995) and the Department of Energy (DE-EE0009285).
}
\thanks{Andrea~Gloppen~Johnsen is with Technical University of Denmark (DTU).}
\thanks{Manuscript received July 25, 2024; revised August 26, 2024.}}

\markboth{Journal of \LaTeX\ Class Files,~Vol.~14, No.~8, August~2015}
{Shell \MakeLowercase{\textit{et al.}}: Bare Demo of IEEEtran.cls for IEEE Journals}

\maketitle

\begin{abstract}
As the transition to sustainable power generation progresses, green hydrogen production via electrolysis is expected to gain importance as a means for energy storage and flexible load to complement variable renewable generation.
With the increasing need for cost-effective and efficient hydrogen production, electrolyzer optimization is essential to improve both energy efficiency and profitability. 

This paper analyzes how the efficiency and modular setup of alkaline hydrogen electrolyzers can improve hydrogen output of systems linked to a fluctuating renewable power supply. To explore this, we propose a day-ahead optimal scheduling problem of a hybrid wind and electrolyzer system. The novelty of our approach lies in  modeling the number and capacity of electrolyzer modules, and capturing the modules' impact on the hydrogen production and efficiency. We solve the resulting mixed-integer optimization problem with several different combinations of number of modules, efficiency and operating range parameters, using day-ahead market data from a wind farm generator in the ERCOT system as an input. 
Our results demonstrate that the proposed approach ensures that electrolyzer owners can better optimize the operation of their systems, achieving greater hydrogen production and higher revenue.
Key findings include that as the number of modules in a system with the same overall capacity increases, hydrogen production and revenue increases.

\end{abstract}

\IEEEpeerreviewmaketitle

\section{Introduction}

\subsection{Background}

As the transition to sustainable power generation continues, the use of renewable energy sources is becoming more important. However, renewable energy sources such as wind are highly variable, and fluctuations can result in periods of excess power production. This surplus energy often goes to waste as it is difficult to store. Hydrogen electrolyzers may be used to mitigate this issue by providing energy storage, grid-balancing services, and operational flexibility to renewable energy power plants. During peak renewable energy production, electrolyzers can use excess energy to produce hydrogen. This grid-following capability of electrolyzers may help reduce curtailment, resulting in an increase in both utilized renewable power and overall revenue. 

However, in order for an electrolyzer to best support intermittent renewable sources, it needs to be able to match their capacity and variability. One way that electrolyzer systems achieve scalability is through implementing multi-modular systems (examples of commercially available module sizes \cite{modularization_review} are shown in table \ref{modules_table}. Please note that the minimum operating capacity of an electrolyzer module is typically expressed as a percentage of the maximum capacity).

An important challenge of electrolyzer operations is the scheduling of hydrogen production, especially with the difficulties of highly variable power sources. Additionally, the hydrogen production curve of an electrolyzer is non-linear, which results in varying efficiencies at varying input power levels. As a result, accounting for hydrogen production and efficiency is critical for optimizing an electrolyzer's impact.

\begin{table}[t]
\centering
\caption{Available Commercial Module Sizes}
\label{modules_table} 
\begin{tabular}{| >{\centering\arraybackslash}m{10em} | >{\centering\arraybackslash}m{6.5em} | >{\centering\arraybackslash}m{6.5em} | }
\hline
Manufacturer & Module Max. Capacity [MW] & Module Min. Capacity [$\%$] \\ [0.5ex] 
\hline
Stargate & 1 & 20 \\ \hline
John Cockerill & 5 & - \\ \hline
Sunfire & 10 & 25 \\ \hline
Thyssenkrupp-Nucra & 20 & 10  \\ \hline
\end{tabular}
\vspace{-1em}
\end{table}

\subsection{Literature Review}

Recent research has emphasized the benefits of optimization of electrolyzer systems to improve efficiency and reduce costs. This literature review examines works in this field, summarizes their impacts, and identifies gaps in current research. 

Many methods for modeling hydrogen electrolyzer systems have been explored. One study developed a mathematical model for alkaline electrolyzers, which included a key non-linear relationship between input power and hydrogen production \cite{Sanchez_electrolyzer}. Another study used Mixed-Integer Linear Programming (MILP) optimization modeling to calculate the optimal amount of electrolyzers required to meet demand, although this study focused on multi-electrolyzer systems rather than modular ones \cite{Varela_scheduling}. These optimization models can be computationally intensive, however, further research on optimization modeling strategies for renewable-hydrogen power plants cautions that over-simplifying models for electrolyzer systems may result in missed operational details that are important for scheduling decisions \cite{andreas_paper}. 

The advantages of modular electrolyzer systems, mainly flexibility and scalability, are also being explored. Recent research on multi-electrolyzer designs examined the load balancing capabilities of such a system, but did not explore modular system benefits \cite{Li_hybrid_system}. Another study discussed industrial modular electrolyzer system designs, focusing on increasing the system's maximum capacity without considering the impact of the modules on overall hydrogen production \cite{modularization_review}. Comparison of single and modular Proton Exchange Membrane (PEM) electrolyzers improved utilization of module efficiency, although neither alkaline electrolyzers nor industrial scale conditions were considered \cite{Makhsoos_PEM_comparative}. A separate review discussed a decentralized modular alkaline electrolyzer design adaptable to changes in the grid, however, it did not address the effect of the modular system on efficiency utilization \cite{Henkel}.

Although previous studies have shown significant improvements in electrolyzer system optimization, there is still a lack of understanding on how modular electrolysis configurations can improve operational flexibility when responding to variable renewable energy inputs. In much of the existing literature, the modeling of the hydrogen production curve is simplified, resulting in the loss of critical peaks in efficiency \cite{andreas_paper}, \cite{Sanchez_electrolyzer}, \cite{Rodrigues_virtual_power}, \cite{Varela_scheduling}, \cite{Makhsoos_PEM_comparative}. Additionally, while there are studies analyzing the potential of using modular systems to increase overall capacity, there is an absence of literature discussing the impact of using smaller electrolyzer modules together in place of a larger single electrolyzer.

\subsection{Contributions}

The goal of this paper is to examine the impact of using multiple smaller electrolyzer modules versus a single large module on the electrolyzer's output and flexibility. Specifically, we study how the number of modules impacts total hydrogen output, power consumption and ability to adjust to fluctuating renewable energy production. 

To support this analysis, we extend the optimization model proposed in \cite{andreas_paper} to model multiple electrolyzer modules. 

Similar to \cite{andreas_paper}, we characterize the efficiency of the electrolyzer through a piecewise-linear approximation of the hydrogen production curve.
We integrate electricity prices and dispatch data for a wind power plant in ERCOT \cite{ERCOT} and use typical hydrogen prices in the optimization model.
Our results demonstrate that multiple modules can improve flexibility and profitability of the electrolyzer. However, to realize those additional benefits, it is necessary to use a detailed model of the hydrogen production curve that captures the efficiency peak in the lower operating range.

In summary, the paper (i) extends existing optimization models for electrolyzers to accommodate modeling of multiple electrolyzer modules and (ii) demonstrates that multi-module systems can improve profitability by allowing individual modules to operate closer to the efficiency peak.

\section{Optimal Scheduling of a Hybrid Power Plant}

In this study, we consider a hybrid power plant comprised of a wind farm connected to an alkaline electrolyzer system and the power grid. The power plant may sell the generated power to the grid at the current day-ahead market price or use the power to produce hydrogen in the electrolyzer. In the following section, the modeling of the electrolyzer module is described, before providing details of the mathematical optimization model used to schedule the hybrid power plant. We use bold, lower-case letters to denote optimization variables and upper-case letters to denote parameters.

\subsection{Electrolyzer Modeling}

The non-linear hydrogen production curve of an electrolyzer (Fig. \ref{fig:h_and_p_curves2}, left) depicts the concave relationship between power consumed and hydrogen produced. The efficiency curve (Fig. \ref{fig:h_and_p_curves2}, right) is the ratio of hydrogen produced per unit input power, and peaks around 30$\%$ of the module's maximum input power capacity before decreasing. 

The hydrogen production data points used in the optimization model below are derived from the empirical model of hydrogen production from \cite{Sanchez_electrolyzer}. The efficiency curve is obtained by dividing the hydrogen output by the power input.

\subsection{Mathematical Model Formulation}

The objective is to maximize profit according to (\ref{eq:obj}): 
\begin{equation}
    \max \quad \sum_{t \in T} \sum_{m \in M} \left[ \lambda_t^{\text{DA}} \boldsymbol{p_t^{\textbf{grid}}} + \lambda^h \boldsymbol{h_{t,m}^e} \right] ,\label{eq:obj}
\end{equation}
where $\boldsymbol{p_t^{\textbf{grid}}}$ is the amount of power sold to the grid in [MW], $\lambda_t^{\text{DA}}$ is the day-ahead electricity market price in [USD/MW], $\lambda^h$ is the price of hydrogen in [USD/kg] (assumed to be a single-value constant), and $\boldsymbol{h_{t,m}^e}$ is the amount of hydrogen produced by the electrolyzer in [kg]. Further, $t$ notes the hour and $m$ notes the electrolyzer module. The objective (\ref{eq:obj}) is subject to the following constraints: \\

\subsubsection{Hydrogen Production Curve}
\label{sec:HPC}
Constraint (\ref{eq:c1_hydrogen_curve}) limits the amount of hydrogen produced by an electrolyzer module $m$ at time $t$ to the piecewise-linear approximation of the hydrogen production curve shown below as:
\begin{align}
    &\boldsymbol{h_{t,m}^e} \leq a_i \boldsymbol{p_{t,m}^e} + b_i C^{\text{max}}  &&\forall t \in T, m \in M, i \in I, \label{eq:c1_hydrogen_curve}
\end{align}
where $a_i$, $b_i$ are the slope and intercept coefficients respectively for piece $i$ of the piece-wise linear (PWL) approximation of the hydrogen production curve,  $C^{\text{max}}$ is the maximum operating capacity of the electrolyzer modules in [MW], and $\boldsymbol{p_{t,m}^e}$ is the power consumed by the electrolyzer in order to produce hydrogen. The use of an inequality rather than an equality in this constraint is a relaxation of the PWL approximation.  This allows the model to produce less hydrogen for a given power input, however, the optimal solution is always equal to the maximum allowable hydrogen output as the linear pieces $i \in I$ form a concave hull.\\

\subsubsection{Operating Range} 
Constraint (\ref{eq:c2_operating_range}) limits the electrolyzer module's production to the minimum and maximum capacity when it is active, and to 0 when the module is off:
\begin{align}
    &C^{\min} \boldsymbol{z_{t,m}^{\textbf{ON}}} \leq \boldsymbol{p_{t,m}^e} \leq C^{\max} \boldsymbol{z_{t,m}^{\textbf{ON}}} && \forall t \in T, m \in M \label{eq:c2_operating_range}
\end{align}
where $C^{\min}$ is the minimum operating capacity of the electrolyzer modules in [MW], and $\boldsymbol{z_{t,m}^{\textbf{ON}}}$ is a binary variable representing the on/off state of module $m$. \\

\subsubsection{Economic Constraint on Grid Power Sales}
Constraint \eqref{eq:c8_grid_constraint1} limits the amount of generated power that may be sold to the grid at time $t$ based on the day-ahead market data (see Data Collection and Preparation for details): 
\label{prob1}
\begin{align}
    &\boldsymbol{p_t^{\textbf{grid}}} \leq P_t^{\text{max}} && \forall t \in T, m \in M ,\label{eq:c8_grid_constraint1}
\end{align}
where $P_t^{\text{max}}$ is the maximum power that may be sold to the grid in [MW] at time $t$. $P_t^{\text{max}}$ is determined by parameters from case study data as shown in \eqref{eq:econ_1}. 
\newline

\subsubsection{Ramp Rate}
Constraints \eqref{eq:c3_ramp_rates_1}-\eqref{eq:c3_ramp_rates_2} limit the change in power consumed by the electrolyzer module $m$ at time $t$: 
\begin{subequations} \label{prob3}
\begin{align}
    &\boldsymbol{p_{(t-1),m}^e} - \boldsymbol{p_{t,m}^e} \leq R && \forall t \in T, m \in M ,\label{eq:c3_ramp_rates_1}\\
    &\boldsymbol{p_{(t-1),m}^e} - \boldsymbol{p_{t,m}^e} \geq -R && \forall t \in T, m \in M ,\label{eq:c3_ramp_rates_2}
\end{align}
\end{subequations}
where $R$ is a constant in [MW], which limits the allowed change in power consumed by the electrolyzer at time $t$ based on the power consumed at the previous time step $t-1$ to $R$. 
\newline

\subsubsection{Start-up time}
Constraints \eqref{eq:c4_startup_1}-\eqref{eq:c6_startup_3} implement binary logic that enforces $\boldsymbol{z_{t,m}^{\textbf{SU}}}=1$ when $\boldsymbol{z_{t,m}^{\textbf{ON}}}$ transitions from off to on, indicating that electrolyzer module $m$ is in 'startup' lasting 1 hour,  and $\boldsymbol{z_{t,m}^{\textbf{SU}}}=0$ otherwise. During a module's startup phase, power may not be consumed in order to produce hydrogen. The logic is implemented in the following constraints: 
\begin{subequations} \label{prob2}
\begin{align}
    &\boldsymbol{z_{t,m}^{\textbf{SU}}} \leq (1 - \boldsymbol{z_{(t-1),m}^{\textbf{ON}}}) && \forall t \in T, m \in M, \label{eq:c4_startup_1}\\
    &\boldsymbol{z_{t,m}^{\textbf{SU}}} \leq \boldsymbol{z_{t,m}^{\textbf{ON}}} && \forall t \in T, m \in M, \label{eq:c5_startup_2}\\
    &\boldsymbol{z_{t,m}^{\textbf{SU}}} \geq \boldsymbol{z_{t,m}^{\textbf{ON}}} - \boldsymbol{z_{(t-1),m}^{\textbf{ON}}} \! \! && \forall t \in T, m \in M, \label{eq:c6_startup_3}
\end{align}
\end{subequations}
where $\boldsymbol{z_{t,m}^{\textbf{SU}}}$ is a binary variable that indicates whether or not module $m$ is in startup phase. \\

\subsubsection{Start-up Cost}
Constraint (\ref{eq:c7_startup_cost}) determines the power consumed during the startup phase as shown below: 
\begin{align}
    &\boldsymbol{p_{t,m}^{\textbf{SU}}} = C^{\text{SU}} \boldsymbol{z_{t,m}^{\textbf{SU}}} && \forall t \in T, m \in M, \label{eq:c7_startup_cost}
\end{align}
where $\boldsymbol{p_{t,m}^{\textbf{SU}}}$ is the power consumed because of startup at time $t$ for module $m$ in [MW], and $C^{\text{SU}}$ is the amount of power required for startup in [MWh]. This power cannot be sold to the grid, or used to produce hydrogen. \\

\subsubsection{Power Balance}
Constraint (\ref{eq:c8_grid_constraint}) maintains balance throughout the hybrid power plant system by ensuring that the net power consumed or sold is never greater than the power generated by the wind farm at time $t$, but allowing curtailment of power, as shown below: 
\begin{align}
    &\boldsymbol{p_t^{\textbf{grid}}} + \boldsymbol{p_{t,m}^e} + \boldsymbol{p_{t,m}^{\textbf{SU}}} \leq P_t^{\text{A}} && \forall t \in T, m \in M. \label{eq:c8_grid_constraint}
\end{align}

\subsection{Optimization Variables and Constraints}
The following variables, objective function, and constraints form a mixed-integer linear program (MILP) implemented with the JuMP package \cite{JuMP}, in the Julia programming language \cite{Julia} using the Gurobi solver \cite{Gurobi}.

\begin{align}
    &\max_{\boldsymbol{x,z}} &&  \mbox{Objective \eqref{eq:obj}} \nonumber\\
&\mbox{s.t.: \,\,\,}  
&& \mbox{Electrolyzer constraints:}~\eqref{eq:c1_hydrogen_curve}, \eqref{eq:c2_operating_range}, \eqref{eq:c3_ramp_rates_1}\text{-}\eqref{eq:c3_ramp_rates_2}, \eqref{eq:c4_startup_1} \text{-} \eqref{eq:c6_startup_3} \nonumber \\[-2pt]
&&& \mbox{Economic constraints:}~\eqref{eq:c8_grid_constraint1}, \eqref{eq:c7_startup_cost}  \nonumber\\[-2pt]
&&& \mbox{Power Balance constraint:}~\eqref{eq:c8_grid_constraint}  \nonumber \\
&&& \boldsymbol{x}\geq 0,~~ \boldsymbol{z}\in\{0,1\} \nonumber
\end{align}
where $\boldsymbol{x}=\{\boldsymbol{h_{t,m}^e, p_{t,m}^e, p_{t}^{\textbf{grid}}, p_{t,m}^{\textbf{SU}}}\}$ are  continuous, non-negative variables and $\boldsymbol{z}=\{\boldsymbol{z_{t,m}^{\textbf{ON}}, z_{t,m}^{\textbf{SU}} }\}$ are binary variables.

\subsection{Electrolyzer Specifications} 
Specifications for electrolyzer parameters such as $C^{\text{min}}$, $C^{\text{max}}$, $R$, $C^{\text{SU}}$, and $\lambda^h$ are determined as follows: 

\begin{itemize}
    \item Capacity: Each module has a maximum capacity ($C^{\text{max}}$) specified in each experiment, with a minimum operating capacity set to 10\% of $C^{\text{max}}$, aligning with industry standards to ensure safe operation \cite{andreas_paper}.
    
    \item Ramp Rate: The modules have a ramp rate limit (0.15 $C^{\text{max}}$/hour), enabling responsiveness to variable renewable power without exceeding mechanical limits \cite{ramp_rate}.
    
    \item Start-up Energy and Time: Modules require 1\% of $C^{\text{max}}$ as start-up energy, with a one-hour start-up time to stabilize before hydrogen production begins \cite{andreas_paper}.
    
    \item Hydrogen Market Price: A constant hydrogen price of \$2 per kg is used to assess profitability \cite{hydrogen_cost}.
\end{itemize}

\begin{figure}
    \centering
    \includegraphics[width=0.95\linewidth]{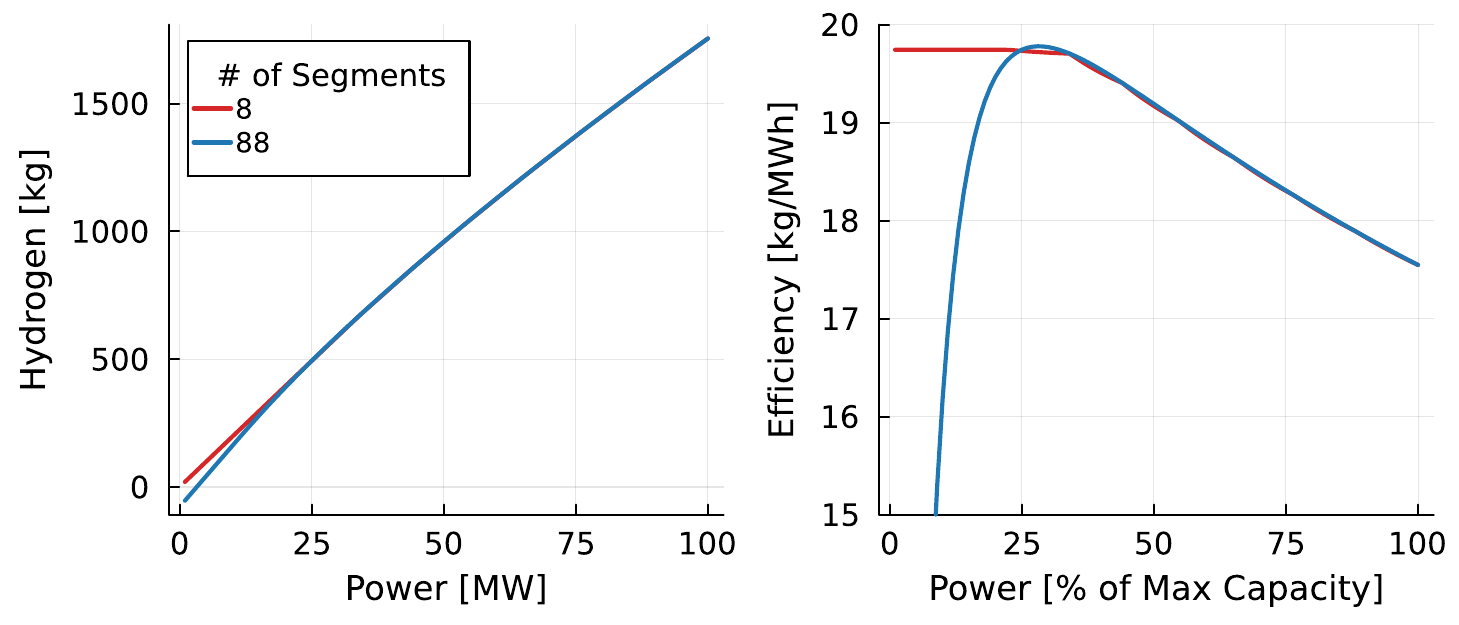} 
    \caption{Hydrogen production (left) and efficiency curves (right) approximated using 8 (blue) or 88 (red)
linear segments}
    \label{fig:h_and_p_curves2}
    \vspace{-1em}
\end{figure}

\section{Numerical Study and Results}

\begin{table*}[t]
\centering
\caption{Performance metrics for systems with different numbers of 88-segment electrolyzer modules over a 1-week period}
\label{tab:performance_metrics}
\begin{tabular}{ccrrrrrr}
\hline
\textbf{\begin{tabular}[c]{@{}c@{}}Number of \\ Modules\end{tabular}} & \textbf{\begin{tabular}[c]{@{}c@{}}Module \\ Capacity \\ {[}MW/h{]}\end{tabular}} & \textbf{\begin{tabular}[c]{@{}c@{}}Total Power \\ Consumed \\ {[}MW{]}\end{tabular}} & \textbf{\begin{tabular}[c]{@{}c@{}}Percent Increase \\ {[}\%{]}\end{tabular}} & \textbf{\begin{tabular}[c]{@{}c@{}}Total Hydrogen \\ Produced \\ {[}kg{]}\end{tabular}} & \textbf{\begin{tabular}[c]{@{}c@{}}Percent Increase \\ {[}\%{]}\end{tabular}} & \textbf{\begin{tabular}[c]{@{}c@{}}Total Revenue \\ {[}USD{]}\end{tabular}} & \textbf{\begin{tabular}[c]{@{}c@{}}Percent Increase \\ {[}\%{]}\end{tabular}} \\ \hline
1 & 100 & 378.80 & 0 & 7408.77 & 0 & 17369.00 & 0 \\
2 & 50 & 400.90 & 5.83 & 7522.56 & 1.54 & 18380.15 & 5.82 \\
4 & 25 & 404.01 & 6.68 & 7646.66 & 3.21 & 18667.75 & 7.48 \\
10 & 10 & 407.20 & 7.50 & 7724.45 & 4.26 & 188823.34 & 8.37 \\ \hline
\end{tabular}
\vspace{-2em}
\end{table*}

\begin{figure}
    \centering
    \includegraphics[width=0.9\linewidth]{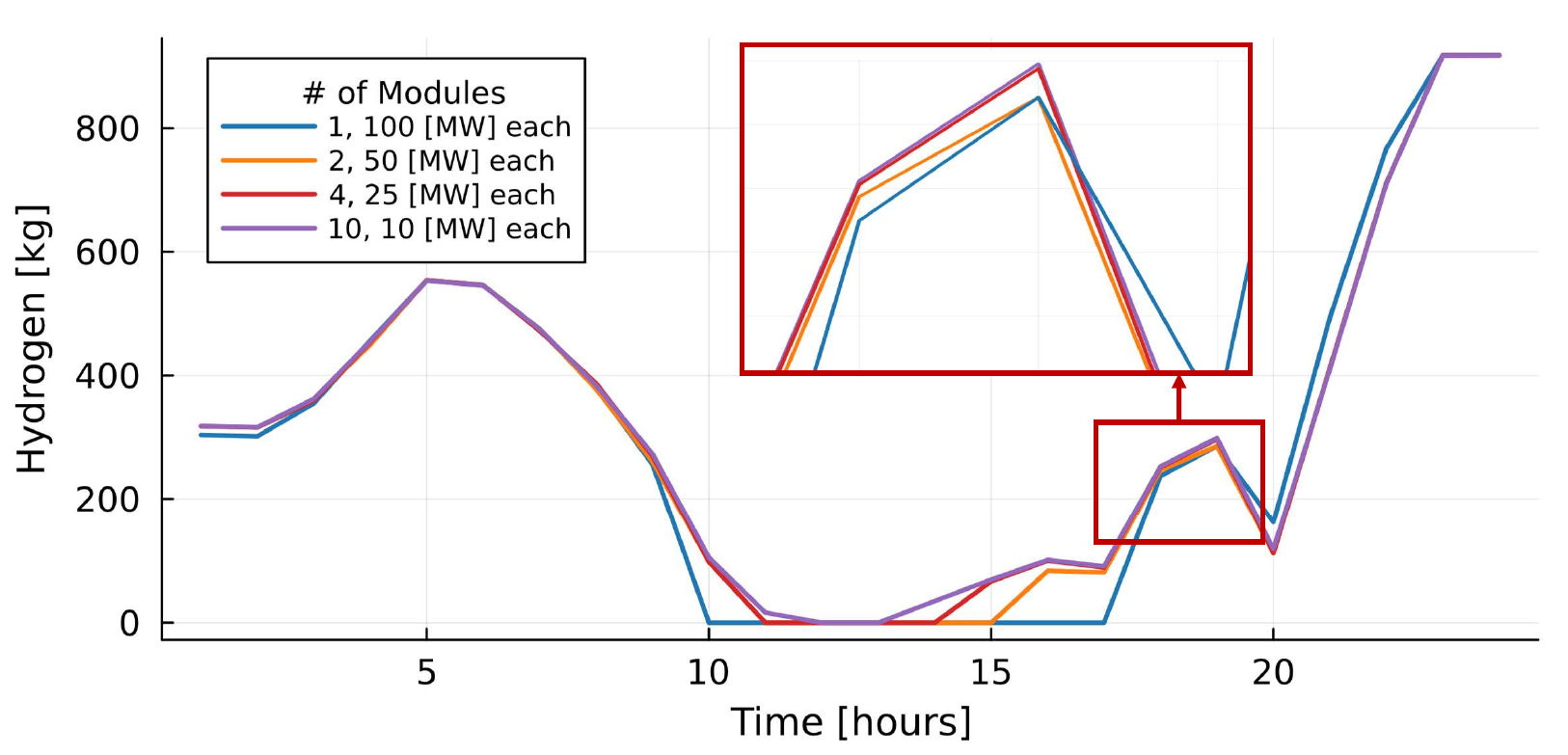}
    \caption{Total hydrogen produced by each multi-modular electrolyzer system over 24 hours, using the 88-segment hydrogen production approximation.}
    \label{fig:h_vs_t_total_88_zoomed}
    \vspace{-0.5em}
\end{figure}

This analysis aims to determine if using multiple smaller electrolyzer modules instead of a single large module can increase the hydrogen output and operational flexibility of an electrolyzer. The analysis was conducted by modeling hydrogen production and efficiency curves (Fig. \ref{fig:h_and_p_curves2}) and comparing performance across different electrolyzer configurations with different numbers of modules (one 100 MW module, two 50 MW modules, four 25 MW modules, and ten 10 MW modules) while keeping the total capacity constant (figures \ref{fig:p_vs_t_total_zoomed}-\ref{fig:wind_p_vs_t_per_module}). No adjustments are made for any ex-post realization of hydrogen production from the given schedule. Performance metrics for each configuration were compared over a one-week period, April 12–18, 2024, capturing total hydrogen produced, power consumed, and profit (see table \ref{tab:performance_metrics}). 

\subsection{Data Collection and Preparation}
Data from ERCOT’s day-ahead market (DAM) was collected for a wind farm near Snyder, Texas, for April 12–18, 2024 \cite{ERCOT}. Key data points include hourly bid prices, cleared prices, high (HSL) and low (LSL) sustainable power limits, and cleared power. These inputs were used to set maximum possible power export limit $P_t^{\text{max}}$ as stated in \eqref{eq:econ_1}. 
\begin{equation}
    P_t^{\text{max}} = 
\begin{cases} 
P_t^{\text{A}} & \text{bid price} > \text{clearing price}, \\ 
P_t^{\text{cleared}} & \text{bid price} \leq \text{clearing price}
\end{cases} \label{eq:econ_1}
\end{equation}

$P_t^{\text{max}}$, applied in  \eqref{eq:c8_grid_constraint1}, is set to the maximum power generated by the wind farm, $P_t^{\text{A}}$, when the bid price is greater than the cleared market nodal price, assuming this indicates that the grid can accept the full bid amount. If the generator bid price is less than market cleared price, we assume congestion limits the sale of additional power to the grid, and the maximum energy sold to the grid is limited to the cleared amount $P_t^{\text{cleared}}$.  These equations use available data from a wind farm to represent the congestion at the location, where the difference between $P^A_t$ and $P^{cleared}_t$ would become curtailed, but here it can be used to power the electrolyzer. Eq. \eqref{eq:econ_1} is implemented ex-ante.

\subsection{Efficiency Curve Approximation}

The hydrogen production and efficiency curves are approximated using two models: an 8-segment and a more detailed 88-segment model (Fig.  \ref{fig:h_and_p_curves2}). The curves are approximated using a piecewise-linear model as posed in Section \ref{sec:HPC}. The coefficients were determined using 100 data points of hydrogen produced at different input power levels \cite{andreas_paper}. The 88-segment model creates a finer approximation of peak efficiency, enabling precise optimization of hydrogen production, especially near power levels associated with peak efficiency. While the 8-segment model offers computational efficiency, it fails to capture peak efficiency accurately.

\subsection{Impact of Multiple Modules}

\begin{figure}
    \centering
\includegraphics[width=0.90\linewidth]{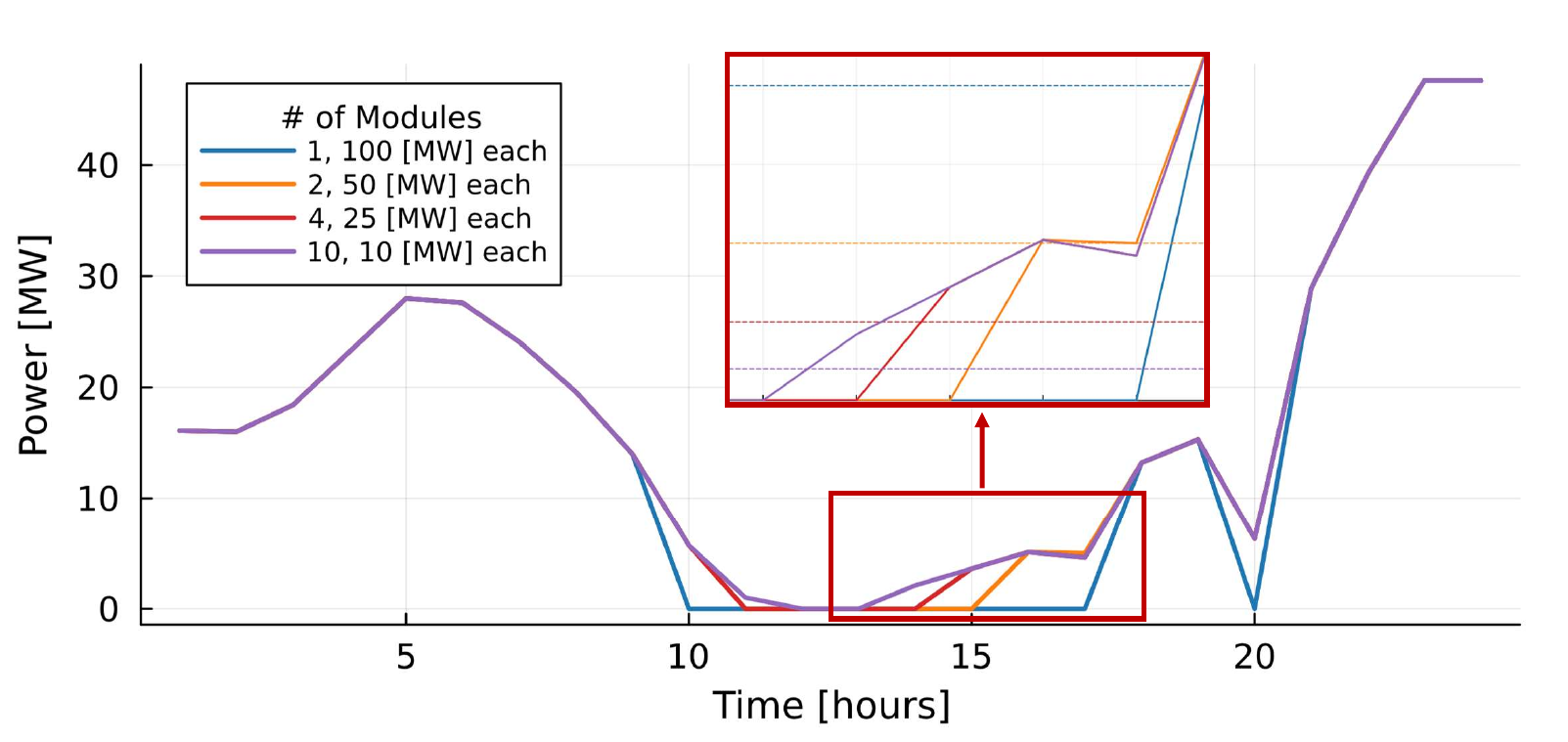}
    \caption{Total wind power consumed by each multi-modular electrolyzer system across a 24-hour period. The lines in the magnified view indicate the minimum power threshold for each configuration.}
    \label{fig:p_vs_t_total_zoomed}
    \vspace{-1em}
\end{figure}

Table \ref{tab:performance_metrics} shows the ex-ante performance of different numbers of modules for an electrolyzer modeled with an 88-segment hydrogen production curve  over a 1-week period. Table \ref{tab:performance_metrics_h18} highlights the differences in schedule between the 8- and 88-segment electrolyzer for a specific hour. Results show that that using multiple modules instead of a single large one significantly increases both hydrogen production and revenue when applying the 88-segment curve. For instance, the configuration with ten 10 MW modules produced 7,724.45 kg of hydrogen and yielded a revenue of $\$$18,882.34 over the study period, representing a 4.26$\%$ increase in hydrogen production and an 8.37$\%$ increase in revenue compared to the single-module system. Two main factors contributed to these gains:

\subsubsection{Lower minimum operating capacity}
Each module's minimum operating capacity ($C^{\text{min}}$) was set to 10$\%$ of its maximum ($C^{\text{max}}$), allowing the multi-module configurations to operate at lower power levels by shutting down some modules. This flexibility increased hydrogen production (Fig. \ref{fig:p_vs_t_total_zoomed}) which is beneficial during times of limited renewable energy availability. Additionally, the ability to operate at lower power levels reduces the need for complete shutdowns, which are necessary when the input power drops below the minimum operating threshold $C^{\text{min}}$ of an individual module. This helps maintain stable hydrogen production
and may help reduce maintenance costs and degradation, though further analysis is needed.
It should be noted that these advantages were observed for both the 8-segment and 88-segment approximations of the hydrogen production curve, indicating that the benefits of reducing $C^{\text{min}}$ are significant for different levels of model complexity. Therefore, it is important to model multiple modules even when the hydrogen production curve is more simply approximated.\\

\subsubsection{Better utilization of the efficiency curve}

\begin{table}[t]
\caption{Performance metrics for different configurations of modules and hydrogen production approximations at hour 18} \label{tab:performance_metrics_h18}
\resizebox{\columnwidth}{!}{
\begin{tabular}{@{}cccccc@{}}
\toprule
\begin{tabular}[c]{@{}c@{}}Number of\\ curve segments\end{tabular} &
  \begin{tabular}[c]{@{}c@{}}Number of\\ Modules\end{tabular} &
  \begin{tabular}[c]{@{}c@{}}Module \\ capacity\\ {[}MW/h{]}\end{tabular} &
  \begin{tabular}[c]{@{}c@{}}Power\\ consumed \\ {[}MW{]}\end{tabular} &
  \begin{tabular}[c]{@{}c@{}}Hydrogen\\ produced \\ {[}kg{]}\end{tabular} &
  \begin{tabular}[c]{@{}c@{}}Profit\\ {[}USD{]}\end{tabular} \\ \midrule
\multirow{4}{*}{8}  & 1  & 100 & 13.2 & 260.6  & 521.21 \\
                    & 2  & 50  & 13.2 & 260.6  & 521.21 \\
                    & 4  & 25  & 13.2 & 260.6  & 521.21 \\
                    & 10 & 10  & 13.2 & 260.6  & 521.21 \\ \midrule
\multirow{4}{*}{88} & 1  & 100 & 13.2 & 237.41 & 474.84 \\
                    & 2  & 50  & 13.2 & 246.79 & 493.59 \\
                    & 4  & 25  & 13.2 & 251.54 & 503.09 \\
                    & 10 & 10  & 13.2 & 253.08 & 506.17 \\ \bottomrule
\end{tabular}
}
\vspace{-1em}
\end{table}

\begin{figure}
    \centering
    \includegraphics[width=0.95\linewidth]{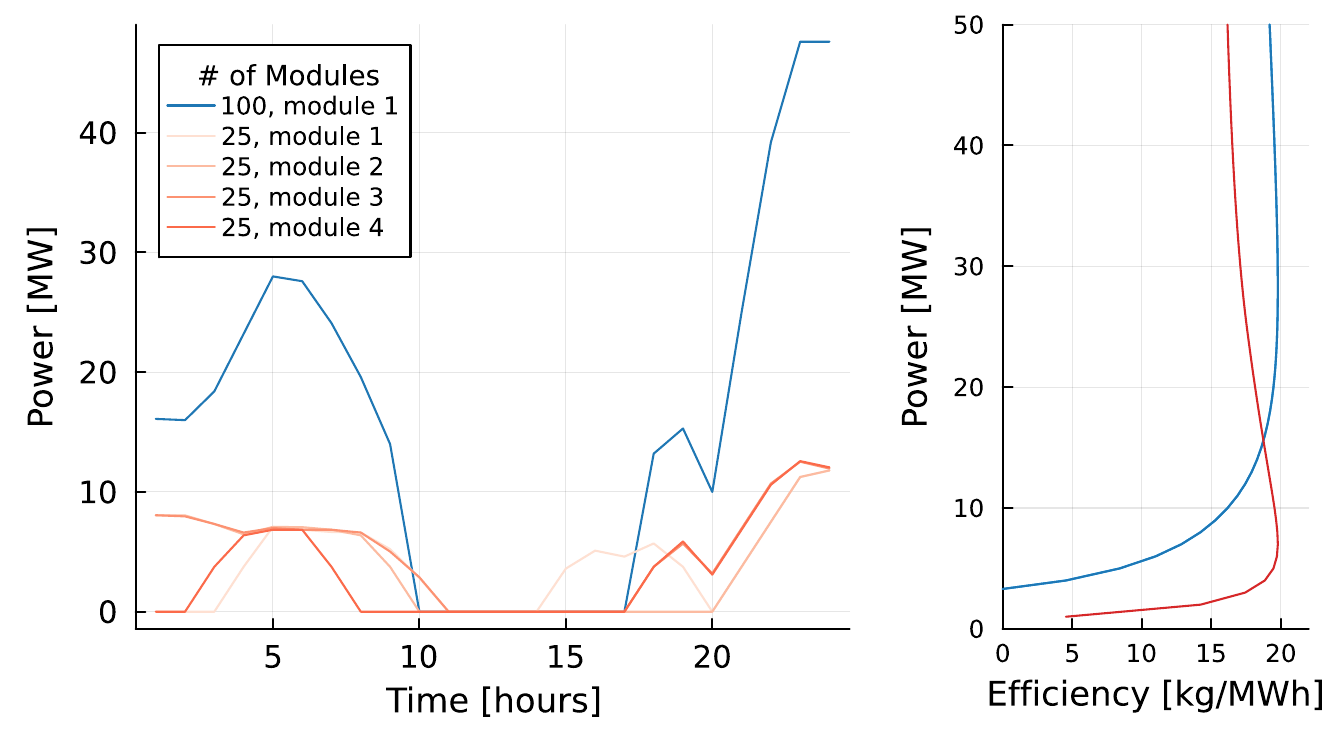}
    \caption{Power consumed by individual modules for 1 (blue) and 4 (red) module electrolyzer systems across 24 hours in alignment with their respective efficiency curves, utilizing the 88-segment hydrogen production approximation.}
    \label{fig:wind_p_vs_t_per_module}
\end{figure}

Multi-module configurations enable the system to operate closer to the efficiency peak by distributing power across modules. This setup allowed more modules to operate closer to peak efficiency, increasing total hydrogen output without increasing energy consumption (Fig. \ref{fig:wind_p_vs_t_per_module}).  However, to fully realize the benefits of this optimized power distribution, the more detailed 88-segment approximation of the hydrogen production curve is necessary to accurately model the performance peak, allowing the optimization algorithm to make better informed decisions about module operation. As shown in table \ref{tab:performance_metrics_h18}, when using the 8-segment approximation at hour 18, each module operates at the same power level rather than varying each module to capture peak efficiency. As a result, increasing the number of modules does not benefit the 8-segment model, whereas when using the 88-segment approximation there is growth in hydrogen production as the number of modules increases because the individual modules set-points are varied to capture the peak efficiency. 
Additionally, no adjustments are made to the schedule for any ex-post realization of hydrogen production, and we expect the 88-segment approximation to more accurately model the production of a real electrolyzer system.

\section{Discussion and Conclusion}

This paper investigated how the use of multiple smaller electrolysis modules rather than a single large module affected system flexibility and hydrogen production. Our findings indicate significant advantages in terms of hydrogen production, revenue, and reduced operating expenses. Specifically, the system's overall minimum operating power $C^{\text{min}}$ is decreased, which permits modules to operate at lower power levels. The flexibility increases operating time, leading to a higher production of hydrogen, as well as reducing the need for complete shutdowns.

Furthermore, by using multiple modules, each module may be more efficiently utilized. The operational parameters of each module can be more optimally controlled, leading to increased hydrogen production and revenue. However, this advantage is only evident when using detailed hydrogen production approximations, which accurately capture the peak efficiency of the electrolyzer module.

The outcomes of this study demonstrate that multi-modular electrolysis systems can enhance the functionality and economics of hydrogen production from intermittent renewable energy sources. Optimizing the control and configuration of these systems can lead to increased system efficiency and profitability. Future research should explore the financial implications and technical challenges associated with implementing multi-module systems, the operational impact on degradation for multi-module systems, as well as investigate the integration of other advanced technologies such as solid oxide electrolysis (SOEC) for high-performance hydrogen production.

\ifCLASSOPTIONcaptionsoff
  \newpage
\fi

LA-UR-24-31641


\end{document}